\documentclass{cimento}
\usepackage{amsmath}

\usepackage{graphicx} 

\title{Light antinuclei in pp collisions at the LHC: production by coalescence and interaction of antinucleons}
\shorttitle{Light antinuclei in pp collisions at the LHC}
\author{N.~Ciavarelli\from{ins:x}\from{ins:y}\ETC,
F.~Ercolessi\from{ins:x}\from{ins:y},F.~Noferini\from{ins:x},F.~Bellini\from{ins:x}\from{ins:y}}
\instlist{\inst{ins:x} INFN, Sezione di Bologna - Bologna, Italy
  \inst{ins:y} Dipartimento di Fisica, Universit\`a di Bologna - Bologna, Italy}

\begin{document}

\maketitle

\begin{abstract}
A unified afterburner framework is presented to describe nucleon--nucleon final-state interactions and production of light (anti)nuclei via coalescence in high-energy pp collisions at the LHC.
The model reproduces qualitatively the measured distributions of light (anti)nuclei without fine-tuning of the model parameters, as well as correlation observables, and can be extended
to beyond proton--proton collisions.
\end{abstract}

\section{Introduction}
The formation mechanism of light nuclei and antinuclei in high-energy collisions
is still not fully understood, yet it is highly relevant for characterising nuclear matter
under extreme conditions~\cite{ref:Bellini} as well as for interpreting potential
dark-matter signals in indirect searches~\cite{ref:vdD}.

In high-energy collisions, nucleons produced by the particle-emitting source interact in the final
state: nucleons that are sufficiently close in phase space and have matching spin configuration at freeze-out can bind to form a nucleus, a process which is modelled by coalescence~\cite{ref:Kapusta80, ref:Butler63}; if they do not bind, the spatial information about the emission source is preserved and can be accessed measuring two-particle momentum correlations, using femtoscopic techniques~\cite{ref:Lisa, ref:ALICE20}.

\section{Coalescence and Correlation Functions}

\subsection{Coalescence}
The probability of two (or more) nucleons to coalesce depends on the pair average transverse momentum \( k_T \), the size of the nucleon source and the size of the nucleus~\cite{ref:Scheibl99,ref:Bellini}.
Experimental observations from ALICE~\cite{ref:ALICE23} at the Large Hadron Collider show that increasing the source size, and thus the average nucleon separation, leads to a decrease in the coalescence probability, reflecting the reduced phase-space density in an
extended emission region.

\subsection{Correlation function}

Final-state interactions and the size of the particle-emitting source can be probed
using femtoscopic techniques derived from Hanbury--Brown--Twiss (HBT) interferometry~\cite{ref:Lisa} through the two-particle correlation function
\(C(k^*)\). This is experimentally defined as \( C(k^*) = \xi\,\frac{S(k^*)}{B(k^*)} \, \),
 where \(k^* = |\vec{p}_1 - \vec{p}_2|/2\) is the relative momentum of the particle pair
in its rest frame. The signal distribution,  \(S(k^*)\), is obtained by pairing particles from the same
event and contains the physical correlations.
The background distribution, \(B(k^*)\), is obtained by pairing particles from different
events and provides an uncorrelated reference sample.
The factor \(\xi\) accounts for the normalisation.

\section{Afterburner model}
In order to model nucleon--nucleon (N-N) interactions (\textit{e.g.} strong and Coulomb potentials) and coalescence at the same time, an afterburner has
been developed. It is a post-processing coalescence-based simulation module operating event-by-event on the final-state particles simulated upstream by the PYTHIA8~\cite{ref:Sjostrand08} Monte Carlo event generator.
Two theoretical formalisms are implemented: one based on wavefunctions and one on
Wigner functions.
At present, the Wigner formalism is primarily used to validate the wavefunction
approach.
The framework extends a previously-developed Wigner-based afterburner~\cite{ref:Mahlein23},
which successfully reproduces experimental data from ALICE.

\subsection{Wavefunction formalism}
The wavefunction formalism is modelled using an interaction potential within a
non-relativistic approximation.
The interaction is based on the nucleon--nucleon potential obtained by equating
the source energy to the final energy at infinity, corresponding to a vanishing
potential:
\( \left\langle \psi_{\text{source}} \middle| K + V \middle| \psi_{\text{source}} \right\rangle =  \frac{k^{*2}}{2m}\; (v \ll c)\)
where \(K\) is the kinetic energy and \(V\) the interaction potential.
The strong interaction is modelled by a square-well potential adjusted to the
deuteron radius,
\( V_{\text{strong}}(r) = -V_{0}\,\Theta(R_{D}-r) \),
with \(V_{0} = 17.4~\text{MeV}\) and \(R_{D} = 3.2~\text{fm}\).
The Coulomb interaction is given by \( V_{\text{Coul}}(r) = \frac{1.44~\text{MeV\,fm}}{r} \).
The source is described by a Gaussian distribution multiplied by a plane wave,
\( \psi_{\text{source}}(\vec{r}) = A\,\exp\!\left(-\frac{r^{2}}{8R_{0}^{2}}\right)\,\exp\!\left(i\,\frac{\vec{k^*}\cdot\vec{r}}{\hbar}\right) \), where \( A \) is the amplitude and \( R_0 \) the source radius.
The coalescence probability is obtained from the projection of the source wave
function onto the deuteron ground state,
\( P_{\text{coal}} =  \left|\left\langle \psi_{\text{source}} \middle| \psi_{\text{deuteron}} \right\rangle\right|^{2}.  \)

\subsection{Validation of the wavefunction model}
The validation of the wavefunction model is based on the Wigner formalism. The source is described by the Wigner function
\( W(\vec r,\vec p) =  \frac{1}{(2\pi\hbar)^{3}}\, \exp\!\left(-\frac{r^{2}}{4R_{0}^{2}}\right)\, \exp\!\left[-4R_{0}^{2} \left(\frac{\vec k^{*}+\vec p}{\hbar}\right)^{2} \right]\), which is Gaussian in both coordinate and momentum space, with inversely proportional relative widths, \(\sigma_{r} = \hbar/\sigma_{k}\).
The deuteron ground state Wigner function, \(W_{\text{deuteron}}(\vec r,\vec p)\), is obtained from the numerical integration of the corresponding wave function.
The coalescence probability is then given by
\( P(k^{*}) = \int W_{\text{deuteron}}(\vec r,\vec p)\, W_{\text{source}}(\vec r,\vec p;\,k^{*})\, d\vec r\, d\vec p \).
Results show consistency between the wavefunction and Wigner function formalisms. The wavefunction approach is preferred because it is computationally more efficient, approximately 33 times faster.

\section{Results}

\subsection{Particle production}
\begin{figure}[htbp!]
\centering
\includegraphics[width=1.\linewidth]{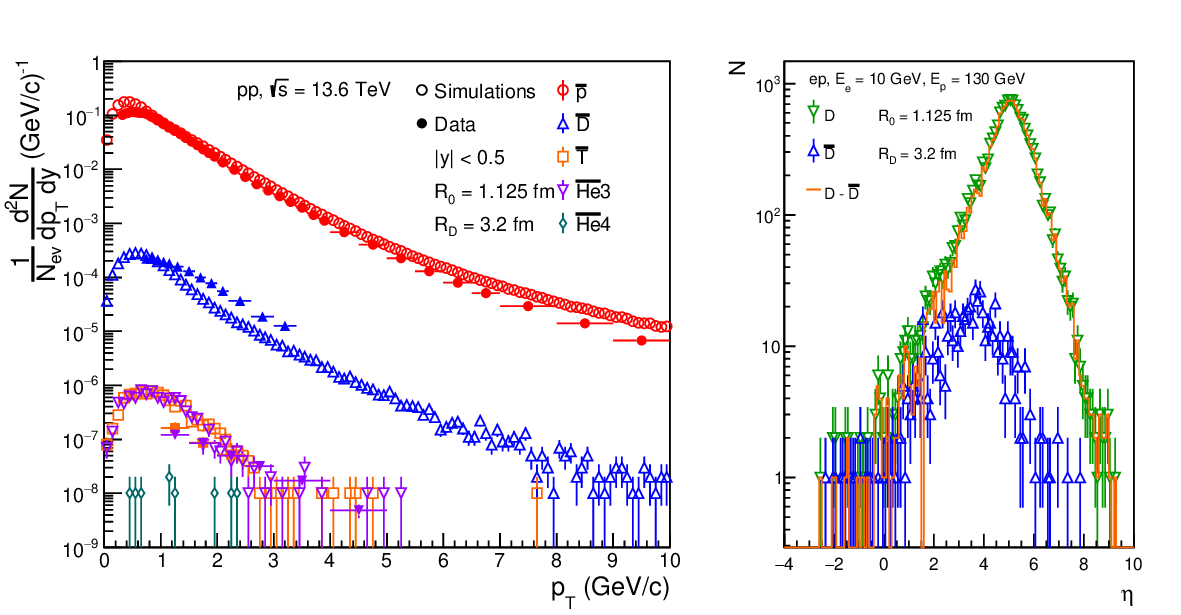}    
\caption{(Left) Transverse momentum distributions for \(\overline{p}\), \(\overline{D}\), \(\overline{T}\), \({}^3\overline{He}\), and \({}^4\overline{He}\) at midrapidity (\(|y| < 0.5\)). Simulated distributions (open markers) are obtained from \(10^{9}\) pp collisions at \(\sqrt{s} = 13.6\)~TeV generated with \textsc{Pythia}, followed by the afterburner, where \(R_{0} = 1.25\)~fm and \(R_{D} = 3.2\)~fm. Data (full markers) are from ALICE~\cite{ref:ALICE21, ref:JHEP22}.
(Right) Pseudorapidity distribution of \(D\), \(\overline{D}\), and their difference for ep collisions simulated with a proton beam energy of 130~GeV and an electron beam energy of 10~GeV, corresponding to an integrated luminosity of $\mathcal{L}=5.33~\mathrm{fb}^{-1}$.}
\label{fig:antiparticleSpec}
\end{figure}

The formation of (anti)nuclei up to \(A = 4\) is simulated by iteratively applying coalescence to the lighter
(anti)nuclei already produced. 
The simulated transverse momentum ($p_T$) distributions of antiproton (\(\overline{p}\)), antideuteron (\(\overline{D}\)), antitriton (\(\overline{T}\)), antihelium-3 (\({}^3\overline{He}\)), antihelium-4 (\({}^4\overline{He}\)), reported in fig.~\ref{fig:antiparticleSpec} (left), are qualitatively consistent with ALICE data \cite{ref:ALICE21, ref:JHEP22}. It should be noted that the excess of (anti)deuterons, as well as other light (anti)nuclei, at low $p_T$ originates from the overestimation of low-$p_T$ (anti)protons in PYTHIA8 rather than from the afterburner.

%In such a model nuclei heavier than \( A > 2 \) are included for the first time.
The model has been applied to predict (anti)deuteron production in electron–proton (ep) collisions at the future Electron Ion Collider~\cite{ref:EICYellow}, 
as shown in fig.~\ref{fig:antiparticleSpec} (right). 
As expected, an asymmetry between \(D\) and \(\overline{D}\) production is observed that favours matter over antimatter.

\subsection{Coalescence in N-N correlations}
\begin{figure}[htbp]
\centering
\includegraphics[width=1.\linewidth]{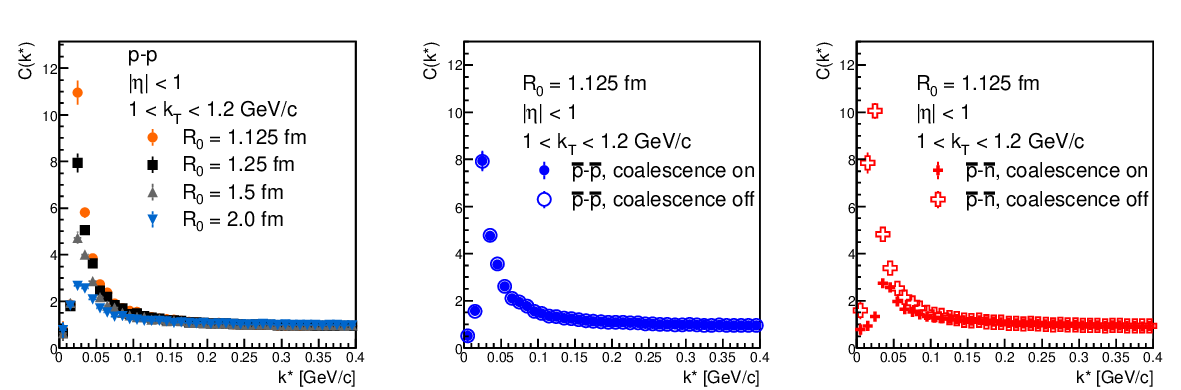}  
\caption{Correlation functions $C(k*)$ of nucleon pairs.
Left: \(\mathrm{p}$–$\mathrm{p}\) correlations for different source radii, $R_0$.
Middle: \(\overline{p}$–$\overline{p}\) correlations with coalescence on and off.
Right: \(\overline{p}$–$\overline{n}\) correlations with coalescence on and off.}
\label{fig:corr}
\end{figure}

Figure~\ref{fig:corr} (left) shows that the afterburner reproduces the effect of
the source size on the correlation function, with the peak of the $C(k*)$ decreasing as the
source radius $R_0$ increases. %,consistent with a more extended spatial emission.
The correlation function for \(\overline{\mathrm{p}}\)--\(\overline{\mathrm{p}}\) pairs remains unchanged by switching coalescence on or off, 
as it is expected since these pairs cannot form bound states (fig.~\ref{fig:corr}, middle).
At very small \(k^*\), the \(C(k^*)\) for p--p and \(\overline{\mathrm{p}}\)--\(\overline{\mathrm{p}}\) pairs is suppressed due to the Coulomb
interaction.
As shown in fig.~\ref{fig:corr} (right), a suppression in the correlation function for \(\overline{\mathrm{p}}\)--\(\overline{\mathrm{n}}\) pairs appears at low \(k^*\) when coalescence is
enabled due to the formation of (anti)deuterons, which removes closely spaced pairs.

\section{Conclusions}
A unified afterburner framework has been developed to model nucleon--nucleon
final-state interactions and light (anti)nuclei production in high-energy
collisions.
The model treats coalescence and correlation effects within a common approach and
has been implemented using both wavefunction and Wigner formalisms, verifying their consistency.
First results with the afterburner show that the framework reproduces qualitatively the
light (anti)nuclei $p_T$ distributions measured at the LHC and captures the effect of bound-state formation on
two-particle correlation functions without fine-tuning of the model parameters.
First predictions for (anti)deuteron production in ep collisions at EIC energies were obtained.
Overall, the afterburner provides a flexible tool that is readily applicable beyond pp collisions, opening new opportunities to systematically investigate nucleon correlations and light (anti)nuclei formation across a wide range of collision systems and energies.


\begin{thebibliography}{0}
\bibitem{ref:Bellini}
\BY{Bellini F. \atque Kalweit A.~P.}
\IN{Phys. Rev. C}{99}{2019}{054905}.

\bibitem{ref:vdD}
\BY{von Doetinchem P.\ \textit{et al.}}
\IN{JCAP}{08}{2020}{035}.

\bibitem{ref:Kapusta80}
\BY{Kapusta J.~I.}
\IN{Phys. Rev. C}{21}{1980}{1301}.

\bibitem{ref:Butler63}
\BY{Butler S.~T. \atque Pearson C.~A.}
\IN{Phys. Rev.}{129}{1963}{836}.

\bibitem{ref:Lisa}
\BY{Lisa M.~A. \atque Pratt S. \atque Soltz R. \atque Wiedemann U.}
\IN{Ann. Rev. Nucl. Part. Sci.}{55}{2005}{357--402}.

\bibitem{ref:ALICE20}
\BY{Acharya S. \etal\ (ALICE Collaboration)}
\IN{Phys. Lett. B}{811}{2020}{135849}.

\bibitem{ref:Scheibl99}
\BY{Scheibl R. \atque Heinz U.~W.}
\IN{Phys. Rev. C}{59}{1999}{1585}.

\bibitem{ref:ALICE23}
\BY{Acharya S.\ \textit{et al.}}
\IN{Phys. Rev. C}{107}{2023}{064904}.

\bibitem{ref:Sjostrand08}
\BY{Sj\"ostrand T. \atque Mrenna S. \atque Skands P.~Z.}
\IN{Comput. Phys. Commun.}{178}{2008}{852}.

\bibitem{ref:Mahlein23}
\BY{Mahlein M. \atque Barioglio L. \atque Bellini F. \atque Fabbietti L. \atque Pinto C. \atque Singh B. \atque Tripathy S.}
\IN{Eur. Phys. J. C}{83}{2023}{804}.

\bibitem{ref:ALICE21}
\BY{Acharya S.\ \textit{et al.}}
\IN{Eur. Phys. J. C}{81}{2021}{256}.

\bibitem{ref:JHEP22}
\BY{Acharya S.\ \textit{et al.}}
\IN{JHEP}{01}{2022}{106}.

\bibitem{ref:EICYellow}
\BY{Abdul Khalek R. \etal}
\IN{Nucl. Phys. A}{1026}{2022}{122447}.



\end{thebibliography}
\end{document}